\documentclass[aps,prd,twocolumn,nofootinbib,nobibnotes,raggedbottom]{revtex4-1}
\pdfoutput=1

\usepackage[usenames,dvipsnames]{xcolor}
\usepackage{hyperref}
\hypersetup{
	colorlinks=true,
	urlcolor=Maroon,
	linkcolor=Maroon,
	citecolor=Maroon,
	bookmarks=true,
	pdftitle={Feynman Integrals and Intersection Theory},
	pdfauthor={Pierpaolo Mastrolia and Sebastian Mizera},
	pdfdisplaydoctitle=true,
	pdfstartview=FitH
}

\usepackage{amsmath}
\usepackage{amssymb}
\usepackage{amsfonts}
\usepackage{mathtools}
\usepackage{microtype}
\usepackage{tikz}
\usepackage{bm}
\usepackage[utf8]{inputenc}
\usepackage{blkarray}
\usepackage{bigstrut}
\usepackage{nccmath}
\usepackage{enumitem}
\usepackage{braket}
\usepackage{dsfont}
\usepackage{lipsum}

\allowdisplaybreaks

\DeclareMathOperator{\Res}{Res}

\def \be {\begin{equation}}
\def \ee {\end{equation}}
\def \nn {\nonumber}
\def \la {\langle}
\def \ra {\rangle}

\def \B {\mathcal{B}}
\def \C {\mathcal{C}}
\def \D {\mathsf{D}}

\def \z {\mathbf{z}}

\numberwithin{equation}{section}

\begin{document}

\title{Feynman Integrals and Intersection Theory}
\author{Pierpaolo Mastrolia}\email{pierpaolo.mastrolia@pd.infn.it}
\affiliation{Dipartimento di Fisica e Astronomia, Universit\`a di Padova, Via Marzolo 8, 35131 Padova, Italy}
\affiliation{INFN, Sezione di Padova, Via Marzolo 8, 35131 Padova, Italy}
\author{Sebastian Mizera}\email{smizera@pitp.ca}
\affiliation{Perimeter Institute for Theoretical Physics, Waterloo, ON N2L 2Y5, Canada}
\affiliation{Department of Physics \& Astronomy, University of Waterloo, Waterloo, ON N2L 3G1, Canada}

\begin{abstract}
We introduce the tools of intersection theory to the study of Feynman integrals, which allows for a new way of projecting integrals onto a basis. In order to illustrate this technique, we consider the Baikov representation of maximal cuts in arbitrary space-time dimension. We introduce a minimal basis of differential forms with logarithmic singularities on the boundaries of the corresponding integration cycles. We give an algorithm for computing a basis decomposition of an arbitrary maximal cut using so-called \emph{intersection numbers} and describe two alternative ways of computing them. Furthermore, we show how to obtain Pfaffian systems of differential equations for the basis integrals using the same technique. All the steps are illustrated on the example of a two-loop non-planar triangle diagram with a massive loop.
\end{abstract}

\maketitle

\section{Introduction}

Evaluating Feynman integrals is crucial for 
the investigation of physical problems that admit  
a field-theoretic perturbative approach: 
revealing weak deviations from the Standard Model predictions in the behaviour of particle collision at high accuracy; the study of properties of newly-discovered particles; understanding formal properties of quantum theories that are not directly deducible from the basic structure of their Lagrangians; exposing similarities among theories which are supposed to describe interactions among particles of different species; as well as the study of the dynamics of coalescing black-hole binary systems whose merger gives rise to gravitational waves. These are just a few examples for which the computation of multi-loop Feynman integrals cannot be considered as optional.

Feynman integrals, within the dimensional regularization scheme, obey contiguity relations known as \emph{integration-by-parts} (IBP) identities \cite{Chetyrkin:1981qh}, which play a crucial role in the evaluation of scattering amplitudes beyond the tree-level approximation.
IBP identities yield the identification of an elementary set of integrals, the so-called \emph{master integrals} (MIs), which can be used as a basis for the decomposition of multi-loop amplitudes.
At the same time, IBP relations can be used to derive differential equations \cite{Barucchi:1973zm,KOTIKOV1991158,KOTIKOV1991123,Bern:1993kr,Remiddi:1997ny,Gehrmann:1999as,Henn:2013pwa,Henn:2014qga}, finite difference equations \cite{Laporta:2001dd,Laporta:2003jz}, and dimensional recurrence relations \cite{Tarasov:1996br,Lee:2009dh} obeyed by MIs. The solutions of those equations are valuable methods for the evaluation of MIs, as alternatives to their direct integration.

IBP identities can be generated by considering integrals of total derivatives that vanish on the integration boundary. They form a system of linear relations between Feynman integrals that differ by the powers of denominators and/or scalar products in the numerator, see, {\it e.g.}, \cite{smirnov2005evaluating,Grozin:2011mt,Zhang:2016kfo}. 
The explicit knowledge of the integration domain 
is not needed for generating IBP identities: 
the requirement of vanishing surface terms provides a sufficient, qualitative information to establish IBP relations. 
Nevertheless, the identification of a basis of MIs, and the integral-decomposition formulas require the solution of large-size linear systems of equations \cite{Laporta:2001dd}. For processes involving multiple kinematic scales, this may represent an insurmountable task. 
In the recent years, important technical advances  have been made by refining the commonly adopted system-solving strategy for IBP identities \cite{Laporta:2001dd}, either due to novel algorithms \cite{vonManteuffel:2014ixa,Peraro:2016wsq,Boehm:2018fpv,Kosower:2018obg,Chawdhry:2018awn} 
or to the development of improved software \cite{vonManteuffel:2012np,Lee:2012cn,Smirnov:2008iw,Maierhoefer:2017hyi,Georgoudis:2016wff}, and 
calculations of multi-loop multi-particle amplitudes, which were considered inaccessible a few years ago, have become feasible  
\cite{Gehrmann:2015bfy,Laporta:2017okg,Badger:2017jhb,Abreu:2018jgq}.

At the same time, one may want to search for alternative methods in order to perform the  decomposition in terms of MIs, eventually looking for mathematical methods that allow for a direct integral reduction, which bypass the need of solving a system of linear equations.

In this work, we explore the latter idea, and we  elaborate on a new method for establishing relations among Feynman integrals in arbitrary space-time dimensions, and for projecting them onto a basis.
An archetype of such a basis reduction is the Gauss's contiguous relation, {\it e.g.},
\be
_2F_1(a,b,c{+}1;z) = \frac{c \, _2F_1(a,b;c;z)}{c-a} + \frac{a \, _2F_1(a{+}1,b;c{+}1;z)}{a-c}.
\ee
This relation can be regarded as a basis reduction of the hypergeometric function $_2F_1(a,b;c{+}1;z)$ on the left-hand side in terms of two MIs on the right-hand side. 
Of course, such an identity can be derived by considering linear relations between $_2F_1$'s with different parameters, derived either from their  series representation or from their integral representation by means of integration-by-parts.
However, it is possible to adopt a modern mathematical approach, which offers a direct solution to the decomposition problem  \cite{aomoto2011theory}. Accordingly,
consider the integral representation,
\be
_2 F_1 (a,b,c{+}1;z) := \frac{1}{B(b,c{-}b)}\!\int_{\mathcal{C}} x^b (1-x)^{c-b} (1{-}zx)^{-a}\, \varphi .
\ee
The integration domain $\mathcal{C} := \overrightarrow{(0,1)}$ together with the information about the branch of the integrand is called the \emph{twisted cycle}, while the single-valued differential form $\varphi := \frac{c}{c-b} d\!\log x$ is called the \emph{twisted cocycle}.\footnote{Strictly speaking, cycles and cocycles are equivalence classes, and here we consider their representatives.} The above integral is understood as a pairing of these two objects \cite{aomoto2011theory}. $B(a,b)$ is the Euler beta function. Similarly, we can consider two other logarithmic forms,
\be
\varphi_1 := d\!\log \frac{x}{1{-}x}, \qquad \varphi_2 := \frac{c}{c{-}b} d\! \log \frac{x}{1{-}zx},
\ee
which upon integration give rise to $_2 F_1(a,b;c;z)$ and $_2 F_1(a{+}1,b;c{+}1;z)$ respectively. The decomposition problem reduces to projecting $\varphi$ onto a basis of $\varphi_1$ and $\varphi_2$. The solution can be found by computing certain topological invariants called \emph{intersection numbers} of pairs of cocycles, which are rational functions in the coefficients $a,b,c$.
They become the coefficients of the basis expansion on the right-hand side of the Gauss' contiguous relation. 

Inspired by this approach, we here propose to apply the computational techniques of {\it intersection theory} to the study of Feynman integrals.

Among different representations of Feynman integrals, the one most closely resembling Aomoto--Gel'fand hypergeometric functions is the so-called \emph{Baikov representation} \cite{Baikov:1996iu}. It uses independent scalar products between external and internal momenta as the integration variables, instead of the components of the loop momenta.
This change of variables introduces a Jacobian equal to the Gram determinant of the scalar products formed by both types of momenta, referred to as the \emph{Baikov polynomial} $\mathcal{B}$, raised to a power $\frac{D-D_\ast}{2}$ for an integer $D_\ast$. Within this representation, it is possible to  identify a critical space-time dimension $D_\ast$, in which the integral presents nice mathematical properties. We review the Baikov representation in Section~\ref{sec:Baikov-representation} and give more details of its derivation in Appendix~\ref{sec:Appendix-A}.

Baikov polynomial fully characterizes the space on which the integrals are defined.
Early signs of this fact were found by Lee and Pomeransky, who applied Morse theory to relate the number of MIs to counting of the critical points of $\mathcal{B}^{\frac{D-D_\ast}{2}}$ \cite{Lee:2013hzt}.\footnote{Similar connections can be made in other representations, such as the parametric one \cite{Lee:2013hzt} (see also \cite{Marcolli:2008vr,marcolli2010feynman}). A correspondence between the Euler characteristic of the sum of Symanzik polynomials and the number of MIs was also recently studied in \cite{Bitoun:2017nre}.} 
IBP identities are non-homogeneous relations which, in general, 
involve integrals associated to a given graph (characterized by a given number of denominators)
and integrals corresponding to subdiagrams
(with fewer denominators). 
The homogeneous terms of IBPs can be detected by maximal cuts, since the simultaneous on-shell conditions annihilate the terms corresponding to subdiagrams. For recent studies of maximal cuts in the Baikov representation, see \cite{Larsen:2015ped,Bosma:2017hrk,Frellesvig:2017aai,Zeng:2017ipr,Bosma:2017ens,Harley:2017qut,Boehm:2018fpv}. In this work, we focus on maximal cuts in order to present our novel algorithm for the basis reduction in the simplest possible setting.

The number of independent MIs can be derived from the properties of Baikov polynomial, therefore, 
after determining the size of the basis, which we denote by $|\chi|$, we construct bases of twisted cycles $\mathcal{C}_i$ and cocycles $\varphi_j$, whose pairings give rise to $|\chi|^2$ integrals:
\be
\int_{\mathcal{C}_i} \mathcal{B}(\z)^{\frac{D-D_\ast}{2}} \varphi_j, \qquad i,j = 1,2,\ldots,|\chi|.
\ee
They form a minimal, \emph{linearly-independent}, basis in terms of which any other integral of the same type can be decomposed. Here $\mathcal{B}(\z)$ is obtained by evaluating $\mathcal{B}$ on the maximal cut surface. Twisted cycles $\mathcal{C}_i$ are chosen as certain regions with boundaries on $\{\mathcal{B}(\z) = 0\}$, while twisted cocycles $\varphi_j$ are differential forms with logarithmic singularities along all the boundaries of the corresponding $\mathcal{C}_j$. We describe them in Section~\ref{sec:minimal-basis}. 
The choice of the independent integrals is rather general, and they might not correspond to the cuts of conventional MIs (although they can be related to them, if needed). 

In Section~\ref{sec:basis-reduction} we apply the tools of intersection theory of the appropriate homology and cohomology groups \cite{MANA:MANA19941660122,cho1995} to the problem of basis reduction. It can be done separately in the space of twisted cycles and cocycles. We focus on the reduction in the space of {\it twisted cocycles} and show how to apply two different techniques of evaluating intersection numbers: 
the special case of logarithmic forms has been treated by one of us in \cite{Mizera:2017rqa}, while the case of general one-forms was discussed by Cho and Matsumoto in \cite{cho1995}. To further support the connection between Feynman calculus and intersection theory, we provide references to the relevant literature for the interested reader.\footnote{Recent applications of the theory of homology and cohomology classes to the coaction of one-loop (cut) Feynman integrals can be found in \cite{Abreu:2017enx}.}

Maximal cuts of MIs obey homogeneous difference and differential equations \cite{Lee:2012te,Remiddi:2016gno,Primo:2016ebd,Primo:2017ipr}. From several studies focusing on the solution of differential equations for Feynman integrals \cite{Laporta:2004rb,Remiddi:2016gno,Primo:2016ebd,Primo:2017ipr,Bosma:2017ens,vonManteuffel:2017hms,Frellesvig:2017aai,Harley:2017qut,Adams:2018kez}, MIs have been identified with the independent components of the integration domain.
Owing to the complete characterization of the integrand and of the integration domain, explicit solutions for the maximal cuts can be found in the Baikov representation. 
In general, MIs obey a system of first-order differential equations, whose corresponding matrix has entries which are rational functions of the kinematic variables and of the dimensional regularization parameter $D$. 
The number of MIs depends on the kinematics of process under consideration and on the number of loops, but there is a certain freedom in choosing them. 
In particular, one may choose a set of MIs whose system of differential equations is {\it linear} in $D$~\cite{Argeri:2014qva}, and use the solutions of the homogeneous equations to write a transformation matrix, known as {\it resolvent matrix} of the homogeneous system (around $D=D_0$, for any chosen value of $D_0$, which can be dictated by the physical dimensions of the problem under consideration)~\cite{Remiddi:2016gno,Primo:2016ebd}. The resolvent matrix is employed to change the basis of MIs, and to define a special set of basic integrals that obey {\it canonical} systems of differential equations~\cite{Henn:2013pwa,Adams:2017tga},
for which the differential equation matrix has a simple $D$-dependent term, factored out of the kinematics. In Section~\ref{sec:Pfaffian-systems}, we discuss Pfaffian systems of differential equations satisfied by the basis integrals, and show that they can be derived using our basis reduction algorithm, without employing the IBP identities.

Even  though  the  techniques  described  here  are  generally applicable, throughout the paper we consider functions admitting integral representations over one variable and leave applications to multi-variate cases until future studies.
For illustration purposes, we apply our novel method to a two-loop non-planar triangle diagram with a massive loop, showing the decomposition algorithm in $D$ dimensions, as well as discussing features of the system of differential equations for the corresponding MIs,
both in $D$ and in $4$ dimensions.

\section{\label{sec:Baikov-representation}Baikov Representation}

Consider scalar Feynman integrals with $L$ loops, $E{+}1$ external momenta, and $N$ propagators in a generic dimension $D$:
\be\label{eq:Feynman-integral}
\mathcal{I}_{\nu_1 \nu_2 \cdots \nu_N} := \int_{\mathbb{R}^D} \prod_{i=1}^{L} \frac{d^D \ell_j}{\pi^{D/2}}  \prod_{a=1}^{N} \frac{1}{\D_a^{\nu_a}}.
\ee
We focus on Euclidean space in all-plus signature for simplicity of discussion. Baikov considered a change of integration variables into all independent scalar products between loop and external momenta \cite{Baikov:1996iu},
\be
q_i \in \{\ell_1, \ell_2, \ldots, \ell_L, p_1, p_2, \ldots, p_E \} ,
\ee 
(here we assume that $D \geq E$, so that external momenta do not satisfy additional relations \cite{Gluza:2010ws,Remiddi:2013joa}). There are $M := LE + \frac{1}{2}L(L+1)$ such kinematic invariants, $\ell_i \cdot q_j$. In order to perform the change of variables, one needs to introduce $M{-}N$ extra inverse denominators $\D_a$ known as \emph{irreducible scalar products} (ISPs) with exponents $\nu_a \leq 0$. The original integral \eqref{eq:Feynman-integral} is recovered when $\nu_{N+1} = \cdots = \nu_M = 0$. After the dust settles, one finds \cite{Lee:2009dh,Lee:2010wea}:
\begin{gather}\label{eq:Baikov-representation}
\mathcal{I}_{\nu_1 \nu_2 \cdots \nu_{M}} = c\! \int_{\Gamma} \B^{\gamma}\, \frac{\prod_{i=1}^{L} \prod_{j=i}^{E+L} d(\ell_i \!\cdot\! q_j)}{\prod_{a=1}^{M} \D_a^{\nu_a}},\\
\gamma :=\!
\frac{D{-}E{-}L{-}1}{2},\nn
\end{gather}
with a constant Jacobian $c$, which we will drop from now on. The integrand involves a \emph{rescaled Baikov polynomial}:
\be\label{eq:Baikov-polynomial}
\B := \frac{\det \mathbf{G}_{\{\ell_1, \ldots, \ell_L, p_1, \ldots, p_E\}} }{\det \mathbf{G}_{\{p_1, \ldots, p_E\}}},
\ee
which is a ratio of two Gram determinants. Recall that a Gram matrix of Lorentz vectors $\{q_i\}$ is defined as $\mathbf{G}_{\{q_i\}} := [q_i \!\cdot\! q_j]$. The integration domain $\Gamma$ is given by imposing $L$ conditions:
\be\label{eq:integration-region}
\Gamma_i := \left\{ \frac{\det \mathbf{G}_{\{\ell_i, \ldots, \ell_L, p_1, \ldots, p_E\}} }{\det \mathbf{G}_{\{\ell_{i+1}, \ldots, \ell_L, p_1, \ldots, p_E\}}} > 0 \right\},
\ee
so that $\Gamma := \Gamma_1 \cap \Gamma_2 \cap \cdots \cap \Gamma_L$. Notice that this implies $\B > 0$ everywhere on the integration domain. For convenience, we review a derivation of the Baikov representation in Appendix~\ref{sec:Appendix-A}.

The representation \eqref{eq:Baikov-representation} is particularly friendly towards computing cuts. By a linear transformation we can make a further change of variables into $z_a := \D_a$ for all $M$ inverse denominators. A single cut corresponds to taking a circular integration contour $\{ |z_a| = \varepsilon \}$, which sets $\D_a$ on-shell. Repeating this procedure $N$ times we obtain a \emph{maximal cut}, which takes the general form \cite{Larsen:2015ped}:
\be\label{eq:maximal-cut}
\mathcal{M}_{\nu_1\nu_2\cdots \nu_M} := \int_{\mathcal{C}} \mathcal{B}(\z)^{\gamma}\, \varphi(\z),
\ee
where we ignored an overall constant Jacobian. The integrand depends on the ISPs
\be
\z := (z_{N+1}, z_{N+2}, \ldots, z_M).
\ee
The Baikov polynomial on the maximal cut $\mathcal{B}(\z)$ is given by setting $z_1 = \cdots = z_N = 0$ in eq.~\eqref{eq:Baikov-polynomial}. Similarly, $\varphi(\z)$ is a differential $(M{-}N)$-form obtained as a result of the residue computation. For example, if the original integral had all propagators undoubled, we have $\varphi(\z) = \prod_{a=N+1}^{M} dz_a / z_a^{\nu_a}$. Finally, the integration domain $\C$ is an intersection of $\Gamma$ with the cut surface $\{z_1 = \cdots = z_N = 0\}$. Whenever this intersection is empty, the maximal cut vanishes and the diagram is reducible.

For one-loop diagrams the maximal cut fully localizes the integral and hence the size of the basis is {\it one}, and the coefficients of the decomposition can be obtained simply by means of residue theorem. Let us consider an interesting example of a two-loop non-planar diagram with internal mass $m$ and $p_1^2 = s$, $p_2^2 = p_3^2 = 0$:
\vspace{-.5em}
\be\label{eq:diagram}
\centering
\raisebox{-.35\height}{\includegraphics[scale=1]{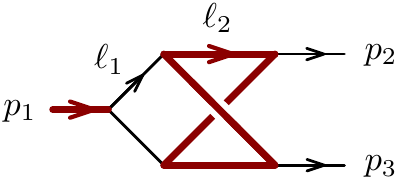}}
\ee
\vspace{-1.5em}
\be
{\begin{aligned}
		&\D_1 = \ell_1^2,\\
		&\D_2 = \ell_2^2 - m^2,\\
		&\D_3 = (p_1 - \ell_1)^2,
\end{aligned}}\qquad
{\begin{aligned}
		&\D_4 = (p_3 - \ell_1 + \ell_2)^2 - m^2,\\
		&\D_5 = (\ell_1 - \ell_2)^2 - m^2,\\
		&\D_6 = (p_2 - \ell_2)^2 - m^2.
\end{aligned}}\nn
\ee
We also choose an ISP $\D_7 = 2(p_2 + \ell_1)^2 - p_1^2$, for later convenience. In this case, $E=L=2$, $M{-}N=1$ and the maximal cut in Baikov representation \eqref{eq:maximal-cut} becomes:
\be\label{eq:maximal-cut-example}
{\cal M}_{111111|-\nu} = \int_{\cal C} {\cal B}(z)^{\frac{D-5}{2}} z^{\nu} dz,
\ee 
where we relabelled $z_7 \to z$, $\nu_7 \to \nu$ for clarity. The rescaled Baikov polynomial reads:
\be\label{eq:Baikov-example}
{\cal B}(z) = -\frac{1}{64s^2} (z^2- s^2)(z^2 - \rho^2),
\ee
where $\rho := \sqrt{s(s+16m^2)}$. In the kinematic regime $s,m^2 >0$ its roots are ordered as $-\rho  < -s < s < \rho$. The constraints \eqref{eq:integration-region} imply that the integration region is:
\be\label{eq:integration-domain}
\mathcal{C} = \overrightarrow{(-\rho, -s)} \cup \overrightarrow{(s,\rho)}.
\ee
Here $\overrightarrow{(a,b)}$ denotes an oriented interval between $a$ and $b$.

\section{\label{sec:minimal-basis}Minimal Basis for Maximal Cuts}

Integrals of the type \eqref{eq:maximal-cut} admit a beautiful interpretation in terms of Aomoto--Gel'fand hypergeometric functions \cite{zbMATH03530308,Gelfand}, where they are understood as pairings of \emph{twisted cycles} $\mathcal{C}$ and \emph{cocycles} $\varphi$. In order to see this, let us consider the following one-form:
\be\label{eq:twist}
\omega := d\log \mathcal{B}(\z)^{\gamma}.
\ee
It defines a flat connection $\nabla_{\omega} := d + \omega \wedge$, related to the integration-by-parts (IBP) relations,
\be
\int_{\cal C} \mathcal{B}(\z)^{\gamma}\, \nabla_{\omega} \xi(\z) = \int_{\cal C} d\left( \mathcal{B}(\z)^{\gamma}\, \xi(\z) \right) = 0
\ee
for any $(M{-}N{-}1)$-form $\xi$ and $(M{-}N)$-dimensional cycle $\C$. This means that we can define equivalence classes $\la \varphi |$ of $(M{-}N)$-forms $\varphi$ up to terms $\nabla_\omega \xi$ that integrate to zero:
\be
\la \varphi | :\quad \varphi \;\sim\; \varphi + \nabla_{\omega} \xi.
\ee
Similarly, we have equivalence classes $|\C]$ of cycles $\C$ up to terms integrating to zero. More precisely, if the integral \eqref{eq:maximal-cut} over the cycle $\widetilde{\C}$ vanishes (for example, when $\widetilde{\C}$ is a loop contractible to a point), then $\C$ and $\C +\widetilde{\C}$ belong to the same equivalence class $|\C]$.\footnote{%
If required, a more rigorous, mathematical definition can be given by following 
Ref.~\cite{aomoto2011theory}:
on the space $X=\mathbb{C}^{M-N} \!\setminus\! \{ {\cal B}(\z) =0 \}$ we define twisted cohomology groups $H^k(X,\nabla_{\omega}) := \ker(\nabla_\omega: \Omega^k(\ast \mathfrak{D}) \to \Omega^{k+1}(\ast \mathfrak{D}))/\nabla_\omega \Omega^{k-1}(\ast \mathfrak{D}) \ni \la \varphi|$, where $\Omega^k(\ast \mathfrak{D})$ is the sheaf of smooth holomorphic $k$-forms on $X$ with poles along the singular divisor $\mathfrak{D}$ of ${\cal B}(\z)$. Locally finite twisted homology groups $H_{k}^{\text{lf}}(X,{\cal L}_\omega) \ni |\C]$ with coefficients in a rank-$1$ local system ${\cal L}_\omega$, are isomorphic by $H^{k}(X,\nabla_{\omega}) \simeq \text{Hom}_{\mathbb{C}}(H_k^{\text{lf}}(X,{\cal L}_\omega),\mathbb{C})$, which induces the pairing $\la \varphi | \C ] := \mathcal{M}_{\nu_1\nu_2\cdots \nu_M}$ in \eqref{eq:maximal-cut}. In this case, $\dim H^k(X,\nabla_{\omega}) = \dim H_{k}^{\text{lf}}(X,{\cal L}_\omega) = \delta_{k,{M-N}}(-1)^{M-N}\chi$.} 
The two classes $\la \varphi |$ and $| \C ]$ encode all IBP identities and contour deformations. Their {\it pairing}, denoted by $\la \varphi | \C]$, is defined to be equal to the maximal cut \eqref{eq:maximal-cut},
\be
\mathcal{M}_{\nu_1\nu_2\cdots \nu_M} = \int_{\mathcal{C}} \mathcal{B}(\z)^{\gamma}\, \varphi(\z) 
=: \la \varphi | \C ] \ .
\ee

In general, Baikov polynomial on the maximal cut ${\cal B}(\z)$ admits a decomposition into irreducible components:
\be\label{eq:Baikov-decomposition}
{\cal B}(\z) = \prod_{i=1}^{K} b_i (\z).
\ee
From now on we will consider only the cases in which each $b_i(\z)$ has degree at most $M{-}N$, so that the corresponding variety $\{ b_i(\z) = 0 \}$ does not have self-intersections. We also assume that the dimension $D$ in the exponent of the Baikov polynomial is \emph{generic}, and in particular not an integer.

The logarithm of ${{\cal B}(\z)}^{\gamma}$ is called a \emph{Morse function} whenever its critical points (also called saddle points) are non-degenerate. From now on we will assume that this is the case. It means that we can use Morse theory to analyze the properties of the integrals \eqref{eq:maximal-cut} and, in particular, to read-off the size of the basis of twisted cycles and cocycles, see, \emph{e.g.}, \cite{milnor2016morse,aomoto2011theory}.

Then the number of independent twisted cycles and cocycles, $|\chi|$, is given by $|\chi| = \sum_{\delta=0}^{M-N} (-1)^{\delta-M+N} C_\delta$, where $C_\delta$ is the number of critical points with a \emph{Morse index} $\delta$. The Morse index is the number of independent directions along which the Morse function decreases away from the critical point \cite{milnor2016morse}. \emph{Under the assumptions} given in \cite{aomoto2011theory}, all critical points have the maximal index $\delta = M{-}N$ and hence:
\be
|\chi| = \big\{ \text{number of solutions of }\omega=0 \big\}.
\ee
Here we used the fact that $\omega = 0$ determines the critical points.\footnote{The idea of determining the size of the basis using the Euler characteristic $\chi$ was first considered in Feynman integral literature by Lee and Pomeransky \cite{Lee:2013hzt}, see also \cite{Bitoun:2017nre}. In the present work, however, we do not make any claims beyond maximal cuts in generic dimension $D$.} When all irreducible components $b_i(\z)$ are linear, $|\chi|$ is also equal to the number of bounded regions in $\mathbb{R}^{M-N} \setminus \{ {\cal B}(\z) = 0 \}$ \cite{aomoto1977structure}.

Let us discuss how to construct bases of twisted cycles $\C_i$ and cocycles $\varphi_i$ for $i=1,2,\ldots,|\chi|$. The real section of the integration domain, $\mathbb{R}^{M-N} \setminus \{ {\cal B}(\z) = 0 \}$, decomposes into multiple disconnected regions, called {\it chambers}. Each chamber is a valid choice of a basis element $\C_i$. For each $\C_i$ we can construct the corresponding twisted cocycle $\varphi_i$ with logarithmic singularities along the boundaries of $\C_i$. For the algorithmic constructions, see, {\it e.g.}, \cite{10006416168,AOMOTO1997119,aomoto2011theory,orlik2013arrangements,KazuhikoAomoto2015} and the more recent study  \cite{Arkani-Hamed:2017tmz}. It is currently known how to do it in the cases when $\C_i$ is bounded by hyperplanes and at most one hypersurface $\{ b_i(\z) = 0 \}$ with degree up to $M{-}N$ \cite{Arkani-Hamed:2017tmz}. This is enough for considering many interesting examples of maximal cuts \cite{Bosma:2017ens}.

For instance, whenever a given chamber $\C_i$ is simplex-like, {\it i.e.}, bounded by exactly $M{-}N{+}1$ hyperplanes $\{ b_j(\z) = 0 \}$, $j=1,2,\ldots,M{-}N{+}1$, we have:
\be\label{eq:simplex-basis}
\varphi_i = d\log \frac{b_1}{b_2} \wedge d\log \frac{b_2}{b_3} \wedge \cdots \wedge d\log \frac{b_{M-N}}{b_{M-N+1}}.
\ee
In situations when $\{ {\cal B}(\z) = 0\}$ is non-normally crossing, {\it i.e.}, more than $M{-}N$ hypersurfaces intersect at a single point, one needs to consider a blowup in the neighbourhood of this point.

The choice of logarithmic basis comes with multiple advantages. For instance, its singularity structure in the variable $\gamma$ is manifest: neighbourhood of each co-dimension $k \leq M{-}N$ boundary of $\C_j$ gives contributions at order $\gamma^{-k}$, {\it e.g.}, hypersurfaces contribute at order $\gamma^{-1}$, while their intersections at order $\gamma^{-2}$:
\be
\centering
\raisebox{-.45\height}{\includegraphics[scale=1]{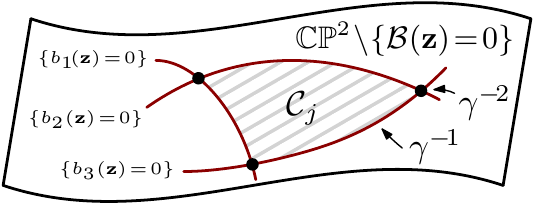}}
\ee
Hence the leading divergence comes from the highest co-dimension boundaries (points), around which the integral behaves as $\gamma^{-(M{-}N)}$. To be precise, a given co-dimension $k$ boundary of the form
\be\label{eq:boundary}
\{ b_1(\z) = 0\} \cap \{ b_2(\z) = 0\} \cap \cdots \cap \{ b_k(\z) = 0\}
\ee
contributes to a given integral $\la \varphi_i | \C_j]$ at order $\gamma^{-k}$ if and only if it belongs to the boundary of a given twisted cycle $\partial\C_j$ and at the same time the twisted cocycle $\varphi_i$ has a non-vanishing residue at the boundary \eqref{eq:boundary}.

At this stage let us remark that it is always possible to shift $D \to D + 2n$ for $n \in \mathbb{Z}$ in the exponent of the Baikov polynomial \cite{Tarasov:1996br}, at a cost at redefining $\varphi \to \mathcal{B}(\z)^{-n} \varphi$ (the requirement of single-valuedness of $\varphi$ imposes $n \in \mathbb{Z}$). This changes $\gamma \to \gamma + n$ and hence the singularity structure, but does not affect the choice of the basis itself.

We can organize all possible pairings of twisted cycles and cocycles into the \emph{twisted period matrix} $\mathbf{P}$ with entries:
\be\label{eq:twisted-period-matrix}
\mathbf{P}_{ij} := \la \varphi_i | \C_j ] = \int_{\C_j} {\cal B}(\z)^{\gamma}\, \varphi_i.
\ee
Its $|\chi|^2$ components provide a minimal linearly-independent basis for any integral of the form \eqref{eq:maximal-cut},
as was first observed in \cite{Primo:2016ebd}, and later also \cite{Frellesvig:2017aai,Zeng:2017ipr,Bosma:2017ens} in the Baikov representation. 
In other words, 
by choosing $\varphi_i$ as basic integrands for  the maximal cuts of the master integrals, 
the entries of the ${\bf P}$ matrix correspond to the independent solutions of the homogeneous system of differential equations the latter obey.  
Let us stress that elements of the basis might not be necessarily recognized as Feynman integrals on their own (but they may correspond to non-trivial combination of them).
For example, evaluating \eqref{eq:twist} for the diagram \eqref{eq:diagram} we have:
\be
\omega = (D{-}5) \frac{ z \left(2z^2 - s^2 - \rho^2 \right)}{(z^2 - s^2)(z^2-\rho^2)} dz.
\ee
We find $|\chi|=3$ solutions of the condition $\omega =0$:
\be\label{eq:critical-points}
z_\ast = 0,\; \pm\sqrt{\frac{s^2 + \rho^2}{2}}.
\ee
Their positions are irrelevant for the counting, but they will be used later on for other purposes (the size of the basis is alternatively determined to be $3$ from the dimension of the pure braid group associated to \eqref{eq:twisted-period-matrix}.) There is a certain freedom is choosing the bases of twisted cycles, as it is only required that they have endpoints on the points $\{-\rho,-s,s,\rho,\infty\}$. A natural choice is as follows:
\be\label{eq:cycles}
\C_1 = \overrightarrow{(-\rho, -s)}, \quad
\C_2 = \overrightarrow{(-s, s)},\quad
\C_3 = \overrightarrow{(s,\rho)}.
\ee
The choice of twisted cocycles is also arbitrary, as long as they have poles only at the points $\{-\rho,-s,s,\rho,\infty\}$. Given the above basis of cycles, a natural counterpart is:
\begin{gather}\label{eq:cocycle-basis}
\varphi_1 = d \log \frac{z {+} \rho}{z{+}s},\quad
\varphi_2 = d \log \frac{z{+}s}{z{-}s},\nn\\
\varphi_3 = d \log \frac{z{-}s}{z{-}\rho},
\end{gather}
which are designed to have residues $\pm 1$ on the two endpoints of the corresponding $\C_i$. They are special cases of \eqref{eq:simplex-basis}. The matrix $\mathbf{P}$ consists of nine linearly independent Appell functions $F_1$ that form a basis.

Note a symmetry for the two independent kinematic invariants, $s \to -s$, $\rho \to -\rho$, under which
\begin{gather}
\C_1 \leftrightarrow -\C_3, \quad \C_2 \to -\C_2,\nn\\
\varphi_1 \leftrightarrow -\varphi_3, \quad \varphi_2 \to - \varphi_2,
\end{gather}
while $\mathcal{B}(z)$ is invariant. Therefore the entries of the matrix $\mathbf{P}$ are related as $\mathbf{P}_{ij} \to \mathbf{P}_{4-i,4-j}$ and we end up with a five-dimensional functionally-independent basis.

In the Feynman integral literature it is customary to talk about bases of MIs, which have to consist of Feynman integrals, and hence have a fixed integration domain $\C$ and different powers of ISPs. In this language, the diagram \eqref{eq:diagram} in generic dimension $D$ has $3$ linearly independent MIs ($\la \varphi_i | \C]$ for $i=1,2,3$), and $2$ after imposing the above symmetry, in agreement with \cite{Primo:2016ebd,vonManteuffel:2017hms}.

\section{\label{sec:basis-reduction}Basis Reduction with Intersection Numbers}

The goal of a basis reduction is expressing an arbitrary integral of the form \eqref{eq:maximal-cut} in terms of the $|\chi|^2$ basis functions in $\mathbf{P}$. This can be done separately in the space of cycles and cocycles. In order to do so, we introduce the notion of a {\it metric} on these spaces. Assuming existence of dual spaces $|\varphi\ra$ and $[\C|$, let us consider pairings between their basis elements:
\be\label{eq:matrices-C-H}
\mathbf{C}_{ij} := \la \varphi_i | \varphi_j \ra, \qquad \mathbf{H}_{kl} := [ \C_k | \C_l ].
\ee
These pairings are called \emph{intersection numbers}. Using simple linear algebra, we can decompose an arbitrary twisted cocycle $\la \varphi |$ into a basis of $\la \varphi_i |$ as follows:
\be\label{eq:cocycle-decomposition}
\la \varphi | = \sum_{i,j=1}^{|\chi|} \la \varphi | \varphi_j \ra \,(\mathbf{C}^{-1})_{ji}\, \la \varphi_i |,
\ee
Concerning the decomposition of Feynman integrals in terms of basic integrals, \eqref{eq:cocycle-decomposition} constitutes the first main result of this work, hence we define to be the {\it master decomposition formula}.

Similarly for a twisted cycle $|\C ]$ in a basis of $|\C_l]$:
\be
| \C ] = \sum_{k,l=1}^{|\chi|} | \C_l ] \,(\mathbf{H}^{-1})_{lk}\, [\C_k | \C].
\ee
Here $\la \varphi | \varphi_j \ra \,(\mathbf{C}^{-1})_{ji}$ and $(\mathbf{H}^{-1})_{lk}\, [\C_k | \C]$ are coefficients of the expansions.
Putting these two decompositions together, we find that the original integral $\la \varphi | \C ]$ is expressed in terms of basis functions in $\mathbf{P}$ as follows:
\be\label{eq:basis-reduction}
\la \varphi | \C ] \;= \sum_{i,j,k,l=1}^{|\chi|}\la \varphi | \varphi_j \ra\, (\mathbf{C}^{-1})_{ji}\, \mathbf{P}_{il}\, (\mathbf{H}^{-1})_{lk}\, [\C_k | \C].
\ee
In fact, this statement is completely general and holds for any Feynman integral in arbitrary parametrization, as long as one is able to identify $|\varphi\ra$ and $[\C|$ and their pairings. The advantage of the Baikov representation of maximal cuts is that such identifications can be made, which allows for explicit computations.

For completeness, we define the dual space as equivalence classes $|\varphi\ra: \varphi \sim \varphi + \nabla_{-\omega} \xi$ with the connection $\nabla_{-\omega}$ and similarly for dual twisted cycles $[\C|$.\footnote{The latter is an equivalence class of cycles $[\C|: \; \C \sim \C + \widetilde{\C}$ such that
\be
\int_{\C} \mathcal{B}(\z)^{-\gamma}\, \varphi(\z) = \int_{\C+\widetilde{\C}} \mathcal{B}(\z)^{-\gamma}\, \varphi(\z)
\ee
for any $\varphi(\z)$. Notice the negative sign in the exponent of the Baikov polynomial compared to \eqref{eq:maximal-cut}.
}
With this choice, intersection numbers $[ \C_i | \C_j]$ are trigonometric functions of the dimension $D$ \cite{MANA:MANA19941660122}. They can be computed straightforwardly by considering all the places where $\C_i$ and $\C_j$ intersect geometrically (additional care needs to be taken when boundaries of $\C_i$ and $\C_j$ are non-normally crossing). In the current manuscript, we will not make use of intersection numbers for cycles: there exist numerous ways of evaluating them, and we refer the reader to, {\it e.g.}, \cite{MANA:MANA19941660122,MANA:MANA19941680111,matsumoto1994,Cho1995-2,Matsumoto1998-2,majima2000,OST2003,aomoto2011theory,yoshida2013hypergeometric,doi:10.1142/S0129167X13500948,Yoshiaki-GOTO2015203,goto2015,goto2015b,Keiji-MATSUMOTO2013367,goto2017,Mizera:2017cqs,relative}.  
In the example at hand, the original Baikov integration domain $\C$ from \eqref{eq:integration-domain} already decomposes as:
\be\label{eq:cycle-decomposition-example}
|\C] = |\C_1] + |\C_3].
\ee
and hence no detailed computation is necessary. Examples for other maximal cuts will appear elsewhere.

Let us stress that intersection numbers entering the expression \eqref{eq:cocycle-decomposition} can be computed for any basis, which does not necessarily have to be the logarithmic one introduced in Section~\ref{sec:minimal-basis}. For example, one could construct a basis of maximal cuts with different powers of ISPs, \emph{e.g.}, \eqref{eq:maximal-cut-example} with $\nu = 0,1,2$.

\subsection{Intersection Numbers of Logarithmic Forms}

Similarly, intersection numbers $\la \varphi_i | \varphi_j \ra$ can be evaluated in multiple different ways, see, {\it e.g.}, \cite{cho1995,matsumoto1998,matsumoto1994,Ohara98intersectionnumbers,Matsumoto1998-2,majima2000,OST2003,aomoto2011theory,yoshida2013hypergeometric,doi:10.1142/S0129167X13500948,Yoshiaki-GOTO2015203,goto2015,goto2015b,Keiji-MATSUMOTO2013367,goto2017,Mizera:2017rqa,relative}. They are rational functions in kinematic invariants and the dimension $D$. It was recently found that for logarithmic forms $\varphi_i$ and $\varphi_j$ there exists a formula localizing on the critical points given by $\omega=0$ \cite{Mizera:2017rqa}:
\be\label{eq:intersection-number}
\la \varphi_i | \varphi_j \ra = (-1)^{M-N}\! \int \prod_{a=N+1}^{M} dz_a\, \delta (\omega_a)\; \widehat{\varphi}_i\, \widehat{\varphi}_j.
\ee
Here $\omega_a$ are components of $\omega = \sum_{a=N+1}^{M} \omega_a dz_a$, and $\widehat{\varphi}$ denotes a differential-stripped cocycle $\varphi =: \widehat{\varphi} \prod_{a=N+1}^{M} dz_i$.

Let use apply it to the two-loop example \eqref{eq:diagram}. For simplicity, we are going to choose the same representatives \eqref{eq:cocycle-basis} for cocycle bases of both $\la \varphi_i |$ and $|\varphi_j\ra$. The above formula \eqref{eq:intersection-number} becomes:
\be
\la \varphi_i | \varphi_j \ra = - \sum_{z_\ast} \frac{1}{\partial \widehat{\omega} / \partial z } \; \widehat{\varphi}_i\, \widehat{\varphi}_j \bigg|_{z=z_\ast},
\ee
where the sum goes over the three critical points $z_\ast$ from \eqref{eq:critical-points} and we have a Jacobian $\partial \widehat{\omega} / \partial z$ coming from evaluating the delta function. Performing this computation for every combination of $\la \varphi_i |$ and $| \varphi_j \ra$ gives us the matrix $\mathbf{C}$ from \eqref{eq:matrices-C-H}:
\be\label{eq:C-matrix}
\mathbf{C} = \frac{2}{D{-}5} \begin{bmatrix}
	2  & -1 & 0 \\
	-1 & 2  & -1 \\
	0  & -1 & 2
\end{bmatrix}.
\ee
It is always possible to choose the dual basis $|\varphi_j\ra$ to be orthonormal, {\it i.e.}, such that $\mathbf{C}=\mathds{1}$, which simplifies the decomposition \eqref{eq:cocycle-decomposition}.\footnote{For instance, an orthonormal basis to \eqref{eq:cocycle-basis} is given by:
\begin{gather}
|\varphi_1\rangle = \gamma \, d\log(z{+}\rho),
\quad
|\varphi_2\rangle = -\gamma \, d\log(z{-}s)(z{-}\rho),\nn\\
|\varphi_3\rangle = -\gamma \, d\log(z{-}\rho),
\end{gather}
with $\gamma = (d-5)/2$,
though we will not make use of it in the text.}

\subsection{Intersection Numbers of Non-Logarithmic One-Forms}

In order to complete the decomposition according to~\eqref{eq:cocycle-decomposition}, we ought to compute the remaining intersection numbers $\la \varphi | \varphi_j \ra$. Let us consider the maximal cut \eqref{eq:maximal-cut-example} with no numerators, $\nu =0$, for which we have $\varphi = dz$. This form has a double pole at infinity, which means that we cannot use \eqref{eq:intersection-number}. In this case we employ an alternative formula for general one-forms due to Cho and Matsumoto \cite{cho1995} (see also \cite{matsumoto1998,zbMATH03996010}):
\be\label{eq:Matsumoto-intersection}
\la \varphi | \varphi_j \ra 
= \sum_{p \in \{ \B(\z) = 0 \}} 
\Res_p \Big[ \psi_p \;\varphi_j \Big] \ , 
\ee
where for each $p$ one needs to compute $\psi_p := \nabla_\omega^{-1} \varphi$ around $z{=}p$, {\it i.e.}, one needs to find a unique holomorphic function $\psi_p$ solving
\be\label{eq:psi-p}
\nabla_\omega \psi_p = \varphi \qquad \text{locally near }p.
\ee
For a review of the derivation of \eqref{eq:Matsumoto-intersection}, see, {\it e.g.}, Appendix~A of \cite{Mizera:2017rqa}. 
For our example, it involves a sum over all boundary components $\{ \B(\z) = 0\} = \{ -\rho, -s, s, \rho, \infty \}$. 
Since the connection $\nabla_{\omega}$ decreases the order of the pole by one, it is enough to consider an ansatz
\be
\psi_p = \sum_{\alpha = 1+\text{ord}_p \varphi}^{\infty} \psi_p^{(\alpha)}\, (z-p)^\alpha,
\ee
where $\text{ord}_p \varphi$ denotes the order of the zero of $\varphi$ around $p$, {\it e.g.}, $\text{ord}_p\, dz/(z-p) = -1$. (Analogous expansion in $1/z$ is done when $p=\infty$.) Plugging it into \eqref{eq:psi-p} one can solve for the first few coefficients $\psi_p^{(\alpha)}$ order-by-order in $(z{-}p)$. Notice that $\text{ord}_p \psi_p = \text{ord}_p \varphi +1$. The residues in \eqref{eq:Matsumoto-intersection} are non-zero only if $\psi_p\, \varphi_j$ has at least a simple pole, or in other words:
\be\label{eq:non-vanishing-condition}
\text{ord}_p \varphi + \text{ord}_p \varphi_j \leq -2.
\ee
In our case, the basis elements $\varphi_j$ have at most simple poles around each $p$, while $\varphi = dz$ has only a double pole at infinity. Using the condition \eqref{eq:non-vanishing-condition} we conclude that $p=\infty$ is the only contributing point in \eqref{eq:Matsumoto-intersection}. Hence we compute:
\be
\psi_\infty = \frac{1}{2D{-}9} z + \mathcal{O}(1)
\ee
using the procedure above, but expanding around infinity. Plugging the result into the formula for intersection numbers \eqref{eq:Matsumoto-intersection} we find:\footnote{For logarithmic forms $\varphi_i$ and $\varphi_j$, the formula \eqref{eq:Matsumoto-intersection} reduces to 
\be
\la \varphi_i | \varphi_j \ra 
= \sum_{p \in \{ \B(\z) =0\}} 
\frac{\Res_p \varphi_i \Res_p \varphi_j}{\Res_p \omega} ,
\ee which is an alternative way of obtaining the entries of $\mathbf{C}$ in \eqref{eq:C-matrix}.}
\begin{gather}
\la\varphi | \varphi_1 \ra = \frac{\rho - s}{2D{-}9}, \quad \la\varphi | \varphi_2 \ra =\frac{2s}{2D{-}9},\nn\\
\la\varphi | \varphi_3 \ra = \frac{\rho - s}{2D{-}9}.
\end{gather}
This completes a decomposition of $\la \varphi |$ into a basis, which after evaluating \eqref{eq:cocycle-decomposition} reads:
\be\label{eq:cocycle-basis-example}
\la \varphi | = \frac{D{-}5}{2(2D{-}9)} \bigg[ \rho\, \big( \la \varphi_1| {+} \la \varphi_2| {+} \la \varphi_3| \big) + s\, \la \varphi_2| \bigg].
\ee
The relation in \eqref{eq:cocycle-basis-example} is an IBP identity.
Finally, using the cycle decomposition \eqref{eq:cycle-decomposition-example}, we find:
\begin{align}
\la \varphi | \C ] \!=\! \frac{D{-}5}{2(2D{-}9)} \! \bigg[ \rho\, \big( \mathbf{P}_{11} {+} \mathbf{P}_{21} {+} \mathbf{P}_{31} \big) &+ s\, \mathbf{P}_{21} \\
&+ (\mathbf{P}_{i 1} {\to} \mathbf{P}_{i 3})\bigg].\nn
\end{align}
Recall that the terms 
$\mathbf{P}_{i 3} = \la \varphi_i | \C_3]$ are related to 
$\mathbf{P}_{i 1} = \la \varphi_i | \C_1]$ by the symmetry $s \to -s, \rho \to -\rho$, therefore only three elemetns of the basis need to be computed. The prefactor of $\gamma = (D{-}5)/2$ guarantees that the integral $\la \varphi | \C ]$ is finite when $\gamma \to 0$.

Let us stress that the choice of logarithmic forms as a basis for the maximal cut is not a limitation. It was inspired by the mathematical literature, but  other choices are possible, such as monomial powers, or, more generally, rational functions of the integration variables. 
The computational load and the resulting formulas may depend on the choice of the basis, but the physical results do not.

Also, we remark that the formula for the intersection numbers \eqref{eq:Matsumoto-intersection} is specific to one-forms, though, in general, maximal cuts of Feynman integrals may admit $m$-forms representations, where the integration variables correspond to the ISPs. Intersection numbers for multivariate logarithmic forms were presented in \cite{matsumoto1998}, and its extension to non-logarithmic forms, and application to Feynman integrals will be discussed in a future publication.

\section{\label{sec:Pfaffian-systems}Pfaffian Systems}

The derivation of Pfaffian systems of differential equations follows the same decomposition algorithm. Let us denote with $d'$ a differential on the appropriately chosen space of kinematic invariants $\cal K$. Acting with it on the basis elements, we have:
\begin{eqnarray}
\label{eq:Pfaffian-derivation}
d'\mathbf{P}_{ij} &=& d' \la \varphi_i | \C_j ] = 
\la \Phi_i | \C_j ] \ , 
\end{eqnarray}
with 
\begin{eqnarray}
\la \Phi_i | \C_j ] &:=&  
\la \, (d' + d'\log {\cal B}^{\gamma}) \ \varphi_i \,  | \C_j ] \ .
\end{eqnarray}
We want to bring it into the Pfaffian form:\footnote{Recall that in the definition of $\mathbf{P}_{ij}$ we stripped away Jacobians, {\it i.e.}, set $c\, c' = 1$ in \eqref{eq:Baikov-appendix}. They can be easily restored and will modify the Pfaffian system.}
\be\label{eq:Pfaffian-system}
d' \mathbf{P}_{ij} = \mathbf{\Omega}_{ik} \wedge \mathbf{P}_{kj} \qquad \text{for all }\, j.
\ee
Here $\mathbf{\Omega}$ is a one-form on $\cal K$. In order to find it, it is enough to project each $\Phi_i$ on the right-rand side of \eqref{eq:Pfaffian-derivation} in the basis $\mathbf{P}$, again using \eqref{eq:cocycle-decomposition}. Hence the entries of $\mathbf\Omega$ are computed with intersection numbers:
\be\label{eq:Omega-decomposition}
\mathbf{\Omega}_{ik} = \sum_{l=1}^{|\chi|} \la \Phi_i | \varphi_l \ra\, (\mathbf{C}^{-1})_{lk}.
\ee
We expect that for the logarithmic bases described in this work the matrix $\mathbf{\Omega}$ should be proportional to $\gamma$.
In addition, it can be shown that the leading order in $\gamma$ of $\mathbf{P}$ coincides with the intersection numbers, {\it i.e.},
\be
\mathbf{P}_{ij} = \mathbf{C}_{ij} + \mathcal{O}(\gamma^{-(M-N)+1}),
\ee
if the bases of cocycles are logarithmic.

Let us derive differential equations in the example at hand. We choose ${\cal K} = \{ x \in \mathbb{C} \,|\, x=  \rho/s \}$, which gives:
\begin{gather}
\Phi_1 = \frac{s}{(z+ s x)^2} \left(-1 + 
\gamma \frac{2s^2 x (x-1)}{(z+s)(z- s x)}  \right) dz \wedge d'\!x,\nn\\
\Phi_2 = \gamma \frac{4 s^3 x}{(z^2 - s^2)(z^2 - s^2 x^2)} dz \wedge d'\!x,\label{eq:Phi-forms}\\
\Phi_3 = \frac{s}{(z-s x)^2} \left( -1 + 
\gamma \frac{2s^2 x (x-1)}{(z-s)(z+s x)}\right) dz \wedge d'\!x,\nn
\end{gather}
where $\gamma = (D{-}5)/2$. 
Using the definition \eqref{eq:Matsumoto-intersection}, we evaluate:
\be
\la \Phi_i | \varphi_l \ra 
=:
{\bf {\hat C}}_{il} \ d'\!x  ,
\ee
with
\be
{\bf {\hat C}}
= \begin{bmatrix}
 \frac{7 x^2{+}2 x{-}1}{(x{-}1) x (x{+}1)} & -\frac{2}{x-1} & -\frac{x{-}1}{x (x{+}1)} \\
 -\frac{2}{x{-}1} & \frac{4 x}{(x{-}1) (x{+}1)} & -\dfrac{2}{x{-}1} \\
 -\frac{x{-}1}{x (x{+}1)} & -\frac{2}{x{-}1} & \frac{7 x^2{+}2 x{-}1}{(x{-}1) x (x{+}1)} \\
\end{bmatrix}.
\ee
Notice that singularities in $x$ can only occur when two punctures out of $\{-sx,-s,s,sx,\infty\}$ collide, {\it i.e.}, where $x$ is $-1,0,1$ or $\infty$. The matrix $\mathbf{C}$ was already computed in \eqref{eq:C-matrix}, which after plugging in \eqref{eq:Omega-decomposition} gives the one-form $\mathbf\Omega$:
\be
\mathbf{\Omega} = \gamma\!\!\!\!\!\! \sum_{p\in \{-1,0,1 \}}\!\!\!\!\!\! \mathbf{\Omega}^{(p)} \,d'\!\log (x{-}p),
\ee
with
\begin{gather}
\mathbf{\Omega}^{(-1)} = \begin{bmatrix}
	1  & 0 & -1 \\
	1 & 2  & 1 \\
	-1  & 0 & 1
\end{bmatrix}\!,\qquad \mathbf{\Omega}^{(0)} = \begin{bmatrix}
	1  & 1 & 1 \\
	0 & 0  & 0 \\
	1  & 1 & 1
\end{bmatrix},
\nn\\
\mathbf{\Omega}^{(1)} = \begin{bmatrix}
	2  & 0 & 0 \\
	-1 & 0  & -1 \\
	0  & 0 & 2
\end{bmatrix}. 
\end{gather}
Note the symmetry $\mathbf{\Omega}_{ij} = \mathbf{\Omega}_{4-i,4-j}$, which descents from symmetries of $\mathbf{P}$. This is a Fuchsian system, which can be solved using standard techniques \cite{Henn:2013pwa,Henn:2014qga,Argeri:2014qva,Lee:2014ioa,Papadopoulos:2014lla}.

\pagebreak
The linear system \eqref{eq:Pfaffian-system} can be converted into a higher-order differential equations for each of the $i$'s separately. For fixed $i$, the solutions of such an equation can be expressed in terms of the basis of $\mathbf{P}_{ij}$ for different $j$'s, {\it i.e.}, different solutions correspond to distinct choices of the integration cycles.

\section{\label{sec:four-dimensions}Four Dimensions}

So far we have been working in a generic dimension $D \notin \mathbb{Z}$. In this section we illustrate how to apply the same techniques directly in the strict $D \to 4$ limit for the maximal cut of the diagram \eqref{eq:diagram}. Setting $D{=}4$ we have:
\be
\mathcal{B}(z)^{-1/2} = \left( - \frac{1}{64 s^2} \left(z^2 - s^2\right)\left(z^2 - \rho^2\right) \right)^{-1/2}\!\!\!\!\!\!\!\!\!\!\!,
\ee
which behaves as $z^{-2}$ when $z \to \infty$ (as opposed to $z^{2(D-5)}$ in the generic case). Hence the integral has a trivial monodromy at infinity, which changes its topological properties. We will see that this means the size of the basis $|\chi|$ drops from $3$ to $2$.

In order to be able to consistently use the techniques of the previous sections and make the connection to elliptic curves known from the literature \cite{Primo:2016ebd,vonManteuffel:2017hms,Broedel:2017kkb}, let us redefine
\be
\widetilde{\mathcal{B}}(z)^{-1/2} := \left(\!-\frac{1}{64 s^2} \frac{z^2 {-} s^2}{z^2 {-} \rho^2}\right)^{\!-1/2}\!\!\!\!\!\!\!\!\!\!\!,\qquad \widetilde{\varphi} := \frac{\varphi}{z^2 - \rho^2}.
\ee
This leaves the combination $\mathcal{B}(z)^{-1/2} \varphi = \widetilde{\mathcal{B}}(z)^{-1/2} \widetilde{\varphi}$ invariant, but makes the integral defined on
\be
X := \mathbb{CP}^1 \setminus \{ -\rho, -s, s, \rho \},
\ee
with the point at infinity included (hence we require that $\widetilde{\varphi}$ does not have poles at $z=\infty$). Indeed, the double-cover of $X$ is an elliptic curve branching at the four points $\{ -\rho, -s, s, \rho \}$. For simplicity, we continue to work directly on the base space $X$.

That the size of the basis is $2$ can be counted by solving $\widetilde{\omega} := -\frac{1}{2}d\log \widetilde{\mathcal{B}}(z) = 0$, which has two solutions, $z_\ast = 0,\infty$. Alternatively, it is easily seen that $|\mathcal{C}_1]$ and $|\mathcal{C}_2]$ from \eqref{eq:cycles} define two independent cycles on the elliptic curve, while $|\mathcal{C}_3]$ is homologous to $|\mathcal{C}_1]$. Finally, let us see the same fact from intersection numbers, by choosing the bases of twisted cocycles $\langle \varphi_i |$ and $| \varphi_j \rangle$ as in \eqref{eq:cocycle-basis}. (This might not be the optimal choice for $D{=}4$, but we use it for consistency.) Using the definition \eqref{eq:Matsumoto-intersection} with $\widetilde{\omega}$, but summing over $p \in \{ -\rho, -s, s, \rho\}$ we find:
\be\label{eq:C-matrix-4d}
\widetilde{\mathbf{C}}
= - \frac{2}{3 \rho^3 (s{+}\rho)} \begin{bmatrix}
	1 & \frac{3\rho}{s{-}\rho} & 1 \\
    \frac{3\rho {-} 2s}{s{-}\rho} & -\frac{6\rho}{s{-}\rho} & \frac{3\rho {-} 2s}{s{-}\rho} \\
	1 & \frac{3\rho}{s{-}\rho} & 1 \\
\end{bmatrix} ,
\ee
where
\be
\widetilde{\mathbf{C}}_{ij} := 
\langle \widetilde{\varphi}_i | \widetilde{\varphi}_j \rangle \ .
\ee
The rank of this matrix is $2$, which signals that there are only two linearly-independent twisted cocycles. From here it is also seen that $\langle \varphi_1 |$ and $\langle \varphi_3 |$ are linearly dependent, and hence we choose the basis to consist of $\langle \varphi_1 |$, $\langle \varphi_2 |$ and similarly for the dual cocycles. This means we can no longer exploit the symmetry properties under $s \to -s$, $\rho \to -\rho$.

As in the previous sections, we decompose $\langle\widetilde{\varphi}| = dz/(z^2{-}\rho^2)$ into the basis by computing \eqref{eq:cocycle-decomposition}:
\be\label{eq:cocycle-basis-4d}
\langle \widetilde{\varphi} | = \sum_{i,j=1}^{2} \langle \widetilde{\varphi} | \widetilde{\varphi}_j \rangle \,(\widetilde{\mathbf{C}}^{-1})_{ji} \langle\, \widetilde{\varphi}_i | = \rho \, \langle \widetilde{\varphi}_1 | + \frac{s{+}\rho}{2} \langle \widetilde{\varphi}_2 |,
\ee
where we used:
\be
\langle \widetilde{\varphi} | \widetilde{\varphi}_1 \rangle = \frac{\rho ^2 {-} 2 s^2 {+} 3 \rho  s}{3 \rho ^3 (\rho^2 {-} s^2)}, \quad \langle \widetilde{\varphi} | \widetilde{\varphi}_2 \rangle = - \frac{2 s}{\rho ^2 (\rho^2 {-} s^2)},
\ee
and we used the corresponding $2{\times}2$ minor of $\widetilde{\mathbf{C}}$ from \eqref{eq:C-matrix-4d}. Alternatively, we could have computed that $\langle \varphi_3| = \langle \varphi_1|$ and obtained the same result \eqref{eq:cocycle-basis-4d} from \eqref{eq:cocycle-basis-example}.

Let us derive differential equations for this basis. The forms $\widetilde{\Phi}_i$ from \eqref{eq:Pfaffian-derivation} are related to those in \eqref{eq:Phi-forms} by $\widetilde{\Phi}_i = \Phi_i/(z^2 {-} \rho^2)$ and $\gamma = -1/2$. Hence we have:
\be
\langle \widetilde{\Phi}_i | \widetilde{\varphi}_l \rangle =:
{\bf {\widehat C}}_{il} \ d'\!x  ,
\ee
with
\be
{\bf {\widehat C}} = \frac{1}{s^4 x \left(x^2{-}1\right)} \begin{bmatrix}
\frac{1}{3 x^2} & \frac{2}{x(1{-}x)} \\
-\frac{2}{x(1{+}x)} & \frac{4}{x^2{-}1} \\
\end{bmatrix}  ,
\ee
Plugging into the expression for $\mathbf{\Omega}$ in \eqref{eq:Omega-decomposition} 
(upon replacing $\varphi \to {\widetilde \varphi}$), we find:
\be
\mathbf{\Omega} = \begin{bmatrix}
\dfrac{1{-}2 x}{(x{-}1) x} & -\dfrac{1}{2 x} \\
\dfrac{2}{x^2{-}1} & -\dfrac{1}{x{+}1} \\
\end{bmatrix} d'x.
\ee
This provides a linear system of two coupled differential equations, $d' \mathbf{P}_{ij} = \mathbf{\Omega}_{ik} \wedge \mathbf{P}_{kj}$. We can solve them to obtain second-order decoupled equations for $\mathbf{P}_{1j}$ and $\mathbf{P}_{2j}$ separately:
\begin{align}
&x (x{+}1)(x{-}1)^2\,  \partial_x^2 \mathbf{P}_{1j} \nn\\
&{+} 2 (x{-}1) (2 x^2 {-}1)\,\partial_x \mathbf{P}_{1j} + (2 x^2 {-} 3 x {-} 1)\, \mathbf{P}_{1j} = 0,\\
&\,x \left(x^2{-}1\right) \partial_x^2 \mathbf{P}_{2j} + \left(5 x^2{-}1\right) \partial_x \mathbf{P}_{2j}+3 x\, \mathbf{P}_{2j} = 0.
\end{align}
In each case, a basis for the two solutions is provided by different choices of twisted cycles.

\section{Discussion}

The mathematical structure of scattering amplitudes is richer than what is currently known. The study of geometric aspects related to the decomposition of multi-loop amplitudes in terms of a basis of master integrals, and the recent development of ideas for the evaluation of the latter seem to offer a new perspective on Feynman calculus. 
The exploitation of existing relations between multi-loop integrals is of fundamental importance to minimize the computational load required for the evaluation of scattering amplitudes that, according to the number of involved particles and to their masses, depend on several kinematic scales. Integration-by-parts identities have been playing a fundamental role since their discovery, about forty years ago.

In this work, we introduced the tools of intersection theory, 
borrowed from the theory of Aomoto--Gel'fand hypergeometric functions, to Feynman integrals. 
We identified a correspondence between the forms associated to generalized hypergeometric functions and Feynman integrals in the Baikov representation, and demonstrated the applicability of intersection theory, by discussing choices of bases of twisted cycles and cocycles, 
basis reduction in both spaces, as well as derivation of the differential equations.
We applied it to the special case of a 2-loop non-planar 3-point function with internal massive propagators, both in arbitrary $D$ dimensions and in the four dimensional case, which, as studied in the recent literature, is known to involve elliptic integrals.
In particular, we analyzed maximal cuts of Feynman integrals, which have the property that the twisted cocycles 
have singularities only along the vanishing of Baikov polynomials,
that defines the cut surface.
For non-maximal cuts, by definition, there exist singularities 
of twisted cocycles
that do not coincide with 
the cut surfaces. In these cases, one needs to consider \emph{relative} twisted homology and cohomology groups. The definition of intersection numbers in such cases was discussed in the mathematics literature only very recently \cite{matsumoto2018relative}. It is expected that it can also help with lifting the assumption of genericity of $D$, so that one can study integer dimensions directly in more general cases. We leave these questions for future research.

Even in the realm of maximal cuts, the variety defined by 
the vanishing of the Baikov polynomial,
could become more involved at higher loops or with more kinematic scales, which complicates the determination of relevant bases of twisted cocycles. It would be interesting to study whether there exist consistent changes of variables that simplify the decomposition \eqref{eq:Baikov-decomposition} into linear factors, perhaps at the cost of introducing  different exponents, as is the case for, {\it e.g.}, the Appell function $F_4$ \cite{aomoto2011theory}.

Our analysis can be considered as a preliminary exploration of the possibility of developing a method for the direct decomposition of Feynman amplitudes in terms of a basis of independent integrals by means of projections, where the elements of intersection theory allow 
to define a metric within the space of Feynman integrals.

\begin{acknowledgments}
\noindent
We thank the organizers and participants of the UCLA Bhaumik Institute's ``QCD Meets Gravity 2017'' workshop, where this work was initiated.
S.M. thanks Nima Arkani-Hamed, Ruth Britto, Freddy Cachazo, Alfredo Guevara, Keiji Matsumoto, Oliver Schlotterer, Ellis Ye Yuan, and Yang Zhang for useful comments and discussions. 
P.M. wishes to acknowledge stimulating discussions with Amedeo Primo, Ulrich Schubert and Lorenzo Tancredi on contours and starters, and with Hjalte Frellesvig, Federico Gasparotto, Stefano Laporta, Manoj K. Mandal, Luca Mattiazzi, Giovanni Ossola, Ettore Remiddi, Ray Sameshima, William Torres, and Roberto Volpato.  This research was supported in part by Perimeter Institute for Theoretical Physics, and by the Supporting TAlent in ReSearch at Padova University (UniPD STARS Grant 2017). Research at Perimeter Institute is supported by the Government of Canada through the Department of Innovation, Science and Economic Development Canada and by the Province of Ontario through the Ministry of Research, Innovation and Science.  
\end{acknowledgments}

\appendix

\section{\label{sec:Appendix-A}Derivation of the Baikov Representation}

In this appendix we review the derivation of the Baikov representation \cite{Baikov:1996iu,Lee:2009dh,Lee:2010wea} for a scalar Feynman integral with $L$ loops, $E{+}1$ external momenta, and $N$ propagators:
\be
\mathcal{I}_{\nu_1 \nu_2 \cdots \nu_N} := \int_{\mathbb{R}^D} \prod_{i=1}^{L} \frac{d^D \ell_j}{\pi^{D/2}}  \prod_{a=1}^{N} \frac{1}{\D_a^{\nu_a}}.
\ee
Out of the independent momenta (for simplicity we assume that $D \geq E$, though it is not necessary)
\be
q_i \in \{\ell_1, \ell_2, \ldots, \ell_L, p_1, p_2, \ldots, p_E \}.
\ee
involved in the scattering process, we can construct the Gram matrix $\mathbf{G}_{\{q_i\}} := [q_i \!\cdot\! q_j]$, which has $M := LE + \frac{1}{2}L(L+1)$ independent entries depending on loop momenta, $\ell_i \cdot q_j$. We use them to express each of the $N$ inverse propagators as:
\be\label{eq:inverse-propagators}
\D_a = \sum_{i,j} \mathbf{A}^{ij}_a\, \ell_i \!\cdot\! q_j + m_a^2,
\ee
where the sum goes over $M$ kinematic invariants $\ell_i \cdot q_j$ and $m_a$ denotes the mass of an internal particle. Here $\mathbf{A}$ defines an $N \times M$ matrix, whose rows are labelled by propagators and columns by Lorentz products $\ell_i \!\cdot\! q_j$. For later convenience, we extend $\mathbf{A}$ into an $M \times M$ invertible matrix by introducing additional propagators $\D_{N+1}, \D_{N+2}, \ldots, \D_{M}$ called irreducible scalar products. We choose their powers to be $\nu_a \leq 0$, so that they only appear in numerators.

The Baikov representation makes manifest the propagator structure of a given Feynman integral. In order to do so, we first change the integration variables from $\ell^\mu_i$ to $\ell_i \cdot q_j$. This is done by decomposing each loop momentum $\ell_i = \ell^{\parallel}_i + \ell^{\perp}_i$, into the $(E{+}L{-}i)$-dimensional space spanned by $\{ \ell_1, \ell_2, \ldots, \ell_{i-1}, p_1, p_2, \ldots, p_E\}$ and the orthogonal complement \cite{Lee:2010wea}. (The order in which loop momenta are labelled does not matter for the final result.) 
The integration measure becomes:
\begin{widetext}
\begin{align}\label{eq:Baikov-measure}
\int_{\mathbb{R}^D} \frac{d^{D} \ell_i}{\pi^{D/2}} &= \int_{\mathbb{R}^D} \!\!\frac{d^{E+L-i} \ell_i^{\parallel}\, d^{D-E-L+i} \ell_i^{\perp}}{\pi^{D/2}}\nn\\
&= \frac{\pi^{-E-L+i}}{\Gamma\!\left(\frac{D-E-L+i}{2}\right)} \left(\det \mathbf{G}_{\{\ell_{i+1}, \ldots, \ell_L, p_1, \ldots, p_E\}}\right)^{-\frac{1}{2}} \!\!\int_{\Gamma_i} \prod_{j=i}^{E+L} d\!\left( \ell_i \!\cdot\! q_j \right) \left( \frac{\det \mathbf{G}_{\{\ell_i, \ldots, \ell_L, p_1, \ldots, p_E\}} }{\det \mathbf{G}_{\{\ell_{i+1}, \ldots, \ell_L, p_1, \ldots, p_E\}}} \right)^{\!\!\frac{D-E-L-2+i}{2}}\!\!\!\!\!\!\!\!\!\!\!\!\!\!\!\!\!\!\!\!\!\!\!\!.\qquad\quad
\end{align}
\end{widetext}
Gram determinants arise as volumes of parallelotopes spanned by the relevant momenta and enter as Jacobians for the change of variables. (When $i=L$, the Gram determinant in the denominator is that of only external momenta.) In the second line the orthogonal directions $\ell_i^{\perp}$ were integrated out directly in spherical coordinates, which results in the constraint on the integration domain:
\be
\Gamma_i := \left\{ (\ell_i^\perp)^2 = \frac{\det \mathbf{G}_{\{\ell_i, \ldots, \ell_L, p_1, \ldots, p_E\}} }{\det \mathbf{G}_{\{\ell_{i+1}, \ldots, \ell_L, p_1, \ldots, p_E\}}} > 0 \right\}.
\ee

Applying this decomposition to all $L$ loop momenta, we obtain:
\be
\mathcal{I}_{\nu_1 \nu_2 \cdots \nu_{M}} = c\! \int_{\Gamma} \B^{\frac{D-E-L-1}{2}}\, \frac{\prod_{i=1}^{L} \prod_{j=i}^{E+L} d(\ell_i \!\cdot\! q_j)}{\prod_{a=1}^{M} \D_a^{\nu_a}},
\ee
where
\be
c = \frac{\pi^{\frac{L-M}{2}} \left(\det \mathbf{G}_{\{p_1, \ldots, p_E\}}\right)^{-\frac{L}{2}} }{\prod_{i=1}^{L} \Gamma \!\left( \frac{D-E-L+i}{2}\right)},
\ee
is a constant, the integration domain is $\Gamma := \Gamma_1 \cap \Gamma_2 \cap \cdots \cap \Gamma_L$, and the rescaled Baikov polynomial reads
\be
\B := \frac{\det \mathbf{G}_{\{\ell_1, \ldots, \ell_L, p_1, \ldots, p_E\}} }{\det \mathbf{G}_{\{p_1, \ldots, p_E\}}},
\ee
The constraints on $\Gamma$ imply that $\B > 0$ everywhere on the integration domain.

Since the inverse propagators $\D_a$ are related to $\ell_i \!\cdot\! q_j$ by a linear transformation \eqref{eq:inverse-propagators}, we can make an additional change of variables to express the Feynman integrals in terms of $M$ variables $z_a := \D_a$:
\be\label{eq:Baikov-appendix}
\mathcal{I}_{\nu_1 \nu_2 \cdots \nu_{M}} = c\, c'\! \int_{\Gamma} \B^{\frac{D-E-L-1}{2}}\,  \prod_{a=1}^{M}\frac{dz_a}{z_a^{\nu_a}},
\ee
where $c' = (\det \mathbf{A})^{-1}$ is a constant Jacobian. We set $c\, c' = 1$ for convenience. $\B$ and $\Gamma$ are the same quantities as before, but expressed in terms of the new variables $z_a$. The expression \eqref{eq:Baikov-appendix} is called the Baikov representation.

Maximal cut corresponds to choosing the integration contour:
\be
\circlearrowleft_1 \wedge \circlearrowleft_2 \wedge \cdots \wedge \circlearrowleft_N \wedge\, \mathcal{C},
\ee
where $\circlearrowleft_a \,:= \frac{1}{2\pi i }\{ |z_a| = \varepsilon \}$ and $\mathcal{C}$ is the intersection
\be
\C := \Gamma \cap \{ z_1 = 0\} \cap \{ z_2 = 0\} \cap \cdots \cap \{ z_N = 0\}.
\ee
Evaluating \eqref{eq:Baikov-appendix} on this contour, we obtain the Baikov representation of maximal cuts:
\be
\mathcal{M}_{\nu_1\nu_2\cdots \nu_M} := \int_{\mathcal{C}} \mathcal{B}(\z)^{\frac{D-E-L-1}{2}}\, \varphi(\z),
\ee
where $\z = (z_{N+1}, z_{N+2}, \ldots, z_M)$ denotes the collective integration variable and
\be
\mathcal{B}(\z) = \mathcal{B}\,\big|_{z_1 = \cdots= z_N = 0}.
\ee
The $(M{-}N)$-form $\varphi(\z)$ is a result of the residue computation, {\it i.e.},
\be
\varphi(\z) := \mathcal{B}(\z)^{-\frac{D-E-L-1}{2}} \oint_{\circlearrowleft_1 \wedge \cdots \wedge \circlearrowleft_N}\!\!\!\!\!\!\!\!\!\!\!\!\!\!\!\! \mathcal{B}^{\frac{D-E-L-1}{2}} \prod_{a=1}^{M}\frac{dz_a}{z_a^{\nu_a}}.
\ee
For example, when $\nu_1 = \cdots = \nu_N = 1$, we simply have $\varphi(\z) = \prod_{a=N+1}^{M} dz_a / z_a^{\nu_a}$.

\bibliographystyle{JHEP}
\bibliography{references}

\end{document}